\documentclass[jcp,floatfix,reprint, longbibliography,superscriptaddress]{revtex4-2}
\usepackage{graphicx}
\usepackage{color}
\usepackage{array, multirow}
\usepackage{amsmath}
\usepackage{xcolor} 
\usepackage{hyperref}

\usepackage[margin=0.8in]{geometry} 
\hypersetup{
    colorlinks=true, 
    linktoc=all,     
    linkcolor=blue,  
}
\usepackage{amsfonts}
\makeatletter \renewcommand\@make@capt@title[2]{%
\@ifx@empty\float@link{\@firstofone}{\expandafter\href\expandafter{\float@link}}%
\sffamily{\textbf{#1}}\@caption@fignum@sep#2 }

\hypersetup{colorlinks, 
    linkcolor={blue!75!black!80!yellow},
    citecolor={blue!75!black!80!yellow}, 
    urlcolor={blue!75!black!80!yellow}
    }
\DeclareGraphicsExtensions{.eps}
\begin{document}
\title{Branching Quantum Convolutional Neural Networks} 
\author{Ian MacCormack}
\affiliation{Kadanoff Center for Theoretical Physics, University of Chicago, Chicago, Illinois 60637, USA}
\affiliation{Department of Physics, Princeton University, Princeton, New Jersey 08544, USA}
\affiliation{Aliro Technologies, Inc. Boston, Massachusetts 02135, USA}
\author{Conor Delaney}
\affiliation{Aliro Technologies, Inc. Boston, Massachusetts 02135, USA}
\author{Alexey Galda}
\affiliation{James Franck Institute, University of Chicago, Chicago, Illinois 60637, USA}
\affiliation{Aliro Technologies, Inc. Boston, Massachusetts 02135, USA}
\author{Nidhi Aggarwal}
\affiliation{Aliro Technologies, Inc. Boston, Massachusetts 02135, USA}
\author{Prineha Narang}
\email{prineha@seas.harvard.edu}
\affiliation{John A. Paulson School of Engineering and Applied Sciences, Harvard University, Cambridge, Massachusetts 02138, USA}

\date{\today}

\begin{abstract}
Neural network-based algorithms have garnered considerable attention in condensed matter physics for their ability to learn complex patterns from very high dimensional data sets towards classifying complex long-range patterns of entanglement and correlations in many-body quantum systems. Small-scale quantum computers are already showing potential gains in learning tasks on large quantum and very large classical data sets. A particularly interesting class of algorithms, the quantum convolutional neural networks (QCNN) could learn features of a quantum data set by performing a binary classification task on a nontrivial phase of quantum matter. Inspired by this promise, we present a generalization of QCNN, the ``branching quantum convolutional neural network", or bQCNN, with substantially higher expressibility. A key feature of bQCNN is that it leverages mid-circuit (intermediate) measurement results, realizable on current trapped-ion systems, obtained in pooling layers to determine which sets of parameters will be used in the subsequent convolutional layers of the circuit. This results in a ``branching" structure, which allows for a greater number of trainable variational parameters in a given circuit depth. This is of particular use on current-day NISQ devices, where circuit depth is limited by gate noise. We present an overview of the ansatz structure and scaling, and provide evidence of its enhanced expressibility compared to QCNN. Using artificially-constructed large data sets of training states as a proof-of-concept we demonstrate the existence of training tasks in which bQCNN far outperforms an ordinary QCNN. We provide an explicit example of such a task in the recognition of the transition from a symmetry protected topological (SPT) to a trivial phase induced by multiple, distinct perturbations. Finally, we present future directions where the classical branching structure and increased density of trainable parameters in bQCNN would be particularly valuable.
\end{abstract}

\maketitle

\section{Introduction}
\noindent In recent years, the scope and complexity of tasks that classical machine learning can address has grown considerably\cite{Jordan255}. Concurrently, noisy-intermediate scale quantum (NISQ) devices have become more useful and accessible \cite{2019ApPRv...6b1314B, 2018arXiv180100862P,2019arXiv190513641K}, paving the way for an intersection of these fields. With the exponential growth of the Hilbert space with system size in quantum systems, high-dimensional machine learning tasks serve as an early use-case for near-term quantum computers\cite{Dunjko_2018,doi:10.1098/rspa.2017.0551,PhysRevLett.113.130503}. Recent efforts to realize a quantum advantage in machine learning have inspired the creation of many novel variational ans\"atze \cite{QML_review,farhi2018classification,verdon2019quantum,verdonQGNN,kim2017robust,universal_quantum_control, 2019QS&T....4b4001H,PhysRevLett.122.040504,PhysRevA.101.032308,2019Natur.567..209H} for use in applications such as autoencoders \cite{Wan_2017,Romero_2017} and many-body physics\cite{Carleo602}, quantum chemistry \cite{2017arXiv170605413O, butler2018machine,head2020quantum} and image recognition \cite{2016arXiv160505775M}. These variational ans\"atze contain parameters that are trained through repeated feedback and optimization with a classical computer. That is, the quantum device evaluates the circuit for a particular parameter set, the classical computer adjusts the parameters which are then fed back into the quantum device, and this process is repeated until optimal parameters are found \cite{2016NJPh...18b3023M,farhi2017quantum}. While irrefutable `quantum advantage' has yet to be demonstrated via any particular machine learning task, as NISQ devices continue to improve, it is of fundamental as well as technological interest to explore possible advantages of near-term quantum devices in this context.
\newline
\newline
\noindent A circuit ansatz of particular interest is the Quantum Convolutional Neural Network, introduced by Lukin \emph{et al} \cite{2019NatPh..15.1273C} (QCNN), a quantum analog of classical convolutional neural networks (CNN). The stacking of convolution and pooling layers in classical CNNs allows for larger and larger features to be extracted from the data, eventually resulting in a classification output. In the quantum domain, QCNNs use a similar structure, with convolutional filters replaced by entangling gates and pooling layers replaced by controlled rotations and discarding of qubits. Both types of layers contain trainable parameters, which are optimized using a classical optimizer. As the number of parameters grows the classical optimization step becomes more difficult. However, with sufficiently many parameters, QCNNs are in principle capable of performing interesting classification tasks, such as quantum phase recognition and image recognition \cite{franken2020explorations}. Quantum phase recognition, in particular, remains an active and important subject of research across condensed matter and statistical physics \cite{2017NatSR...7.8823B}, towards creating robust phase diagrams and discovering new many-body quantum phenomena in previously unknown parts of phase diagrams.
\newline
\newline
\noindent Typically a portion of the total number of qubits is discarded at each pooling layer in the QCNN, though they need not be measured beforehand. Instead, we can simply replace the mid-circuit measurements and subsequent classically-controlled rotation gates with quantum controlled rotation gates. This replacement is necessary in the majority of existing quantum devices, which do not currently possess the ability to measure qubits mid-circuit, or to condition subsequent operations on these measurement results. However, some trapped-ion quantum devices \cite{2020arXiv200301293P,2020arXiv200603071E}  
do currently offer this ability, and other architectures are working to introduce it in the near future \cite{ibm_quantum_experience_2020}. With the realization of mid-circuit measurements, it is interesting to ask how NISQ algorithms can benefit from this feature, classical conditioning, and control flow. Extensive work has been done on measurement-based quantum computing \cite{2009arXiv0910.1116B}, which makes maximal use of this approach. However, measurement-based-quantum computing requires very large numbers of qubits to perform nontrivial operations. Therefore, in this \emph{Article} we ask a pertinent and timely question, can we incorporate mid-circuit measurement into variational QML ans\"atze in order to take full advantage of the current capabilities of quantum devices, with their limitations on fidelity and circuit depth?
\newline
\newline
\noindent
In this \emph{Article}, we generalize the structure of QCNNs, heavily using mid-circuit measurement capabilities which gives the circuit a classical branching structure where different mid-circuit measurement outcomes dictate which convolutional layers will subsequently be executed. This allows for a greater number of trainable parameters than standard QCNN in a circuit of a fixed gate depth, and therefore a greater level of expressibility (a notion we define more specifically in Section \ref{exp_section}). This is important for the current generation of quantum devices, in which circuit depth is limited by gate noise. We call this QCNN variant ``branching" QCNN (bQCNN) in reference to the classical branching. By comparing the states generated by random QCNN and bQCNN circuits with states from the Haar distribution, we show that the bQCNN is more expressive than a QCNN of the same width and depth (where we do not count mid-circuit measurements towards circuit depth). By applying reversed, randomly-generated bQCNN circuits to product states, we generate training tasks in which the bQCNN significantly outperforms the standard QCNN. To demonstrate a specific instance of bQCNN's advantage, we then compare the performance of the bQCNN and QCNN in learning to recognize the topological phase transition from the $\mathbb{Z}_2 \times \mathbb{Z}_2$ symmetry protected topological phase induced by two distinct types of perturbations to the parent Hamiltonian. Looking ahead, the successful experimental demonstration of a (b)QCNN for classification of quantum phases on a quantum device will pave the way for development and implementation of novel hybrid quantum
machine learning algorithms, which will likely leverage intermediate measurement capabilities of trapped-ion architectures.

\begin{figure}
    \centering
    \includegraphics[width=8cm]{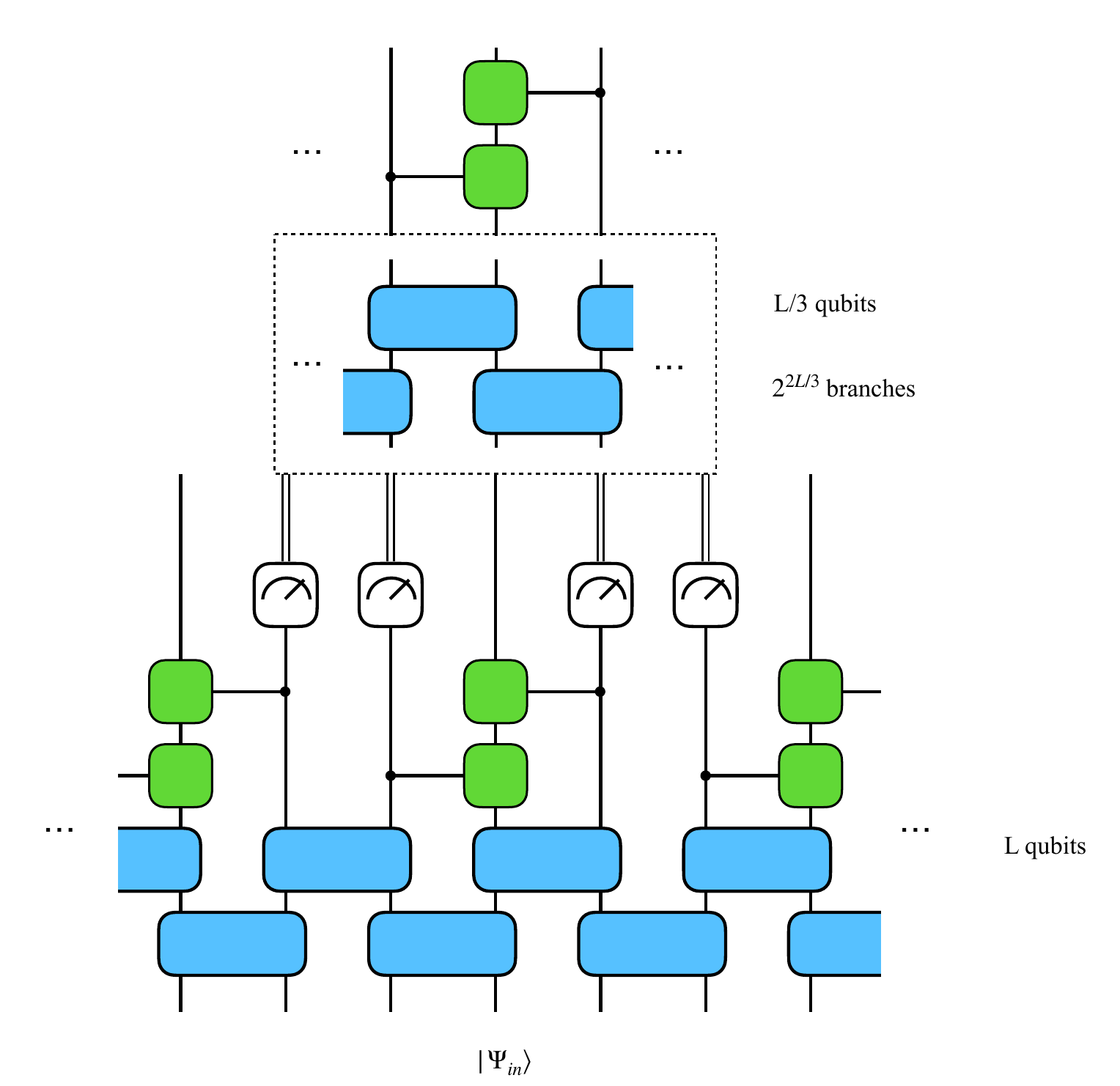}
    \caption{An example of an L-qubit bQCNN. The blue boxes denote parameterized $SU(4)$ 2-qubit gates gates (though one could use many-qubit parameterized gates depending on the problem at hand). The green boxes are controlled single qubit $SU(2)$ operations, and, along with the measurements, constitute the pooling layers. At each pooling layer, the controlled rotations are applied to the qubits that will move onto the next layer, a fraction of the qubits are measured and discarded ($2/3$ of them in this case), And the remaining qubits proceed to the subsequent convolutional layer. \textit{Which} convolutional layer we proceed to (i.e. which branch) is determined by the measurement outcomes of the previous pooling layer. In this example, since we measure $2L/3$ qubits at the first pooling layer, there is a choice of up to $2^{2L/3}$ distinct convolutional layers that can be applied next, each containing its own set of parameterized gates. This pattern of convolution, pooling, branching, convolution is repeated until we are left with a small number of qubits, at which point we execute a small, fully-connected layer of gates, as in \cite{2019NatPh..15.1273C} and measure the designated output qubit(s).}
    \label{fig:IM_vs_nonIM_circuit}
\end{figure}

\section{QCNN vs. bQCNN}
\vspace{-1em}
\subsection{Quantum Convolutional Neural Networks}
\vspace{-1em}
\noindent The quantum convolutional neural network, introduced in \cite{2019NatPh..15.1273C}, is a variational quantum machine learning ansatz that combines unitary convolutional layers with pooling layers. In the pooling layers, a portion of the data qubits is removed and single qubit rotation gates are performed on remaining qubits, conditioned on the states of adjacent qubits. Equivalently, this conditioning can be implemented by performing mid-circuit measurements of the to-be-discarded qubits, with subsequent rotation gates dependent on the measurement outcomes. Convolution and pooling layers are both parameterized by rotation angles that are adjusted by repeated use of a classical optimizer (see Appendix \ref{GA_app} for details on the optimizer we use). The circuit terminates in some number of qubits $N_{out}<N_{in}$, whose measurement outcomes correspond to classifications of the input data. As parameter training progresses, the probability distribution of the output qubits' measurements more closely reproduces the desired classification of the training data.
\newline
\newline
\noindent The QCNN was inspired by the structure of classical convolutional neural networks \cite{lecun2015deep}, where spatial ``filters" are repeatedly applied to data, whose dimensionality is then reduced at a pooling layer. The essential, long-ranged features of the input data are distilled by the combination of the filters and the pooling, eventually resulting in a small number of such features that can be used for a classification or recognition task. The same principles hold in the QCNN. The filters in the convolutional layers are multi-qubit, entangling, parameterizable unitary operators. These serve to disentangle long range correlations in the input state. A pooling layer is then applied to the qubits. Here we select a subset of the qubits to use as controls for controlled rotations that serve to correct single qubit errors (to use language from quantum error correction) in the other qubits. These control qubits are then discarded and the circuit proceeds to the next, smaller convolutional layer. We repeat the convolution, pooling pattern until we are left with a number of qubits suited to provide classifications for the task at hand --- e.g. one qubit for a binary classification task, two qubits for a four category classification task, etc.

\subsection{bQCNN: Leveraging Mid-circuit Measurements}

\begin{figure}
    \centering
    \includegraphics[width=4 cm]{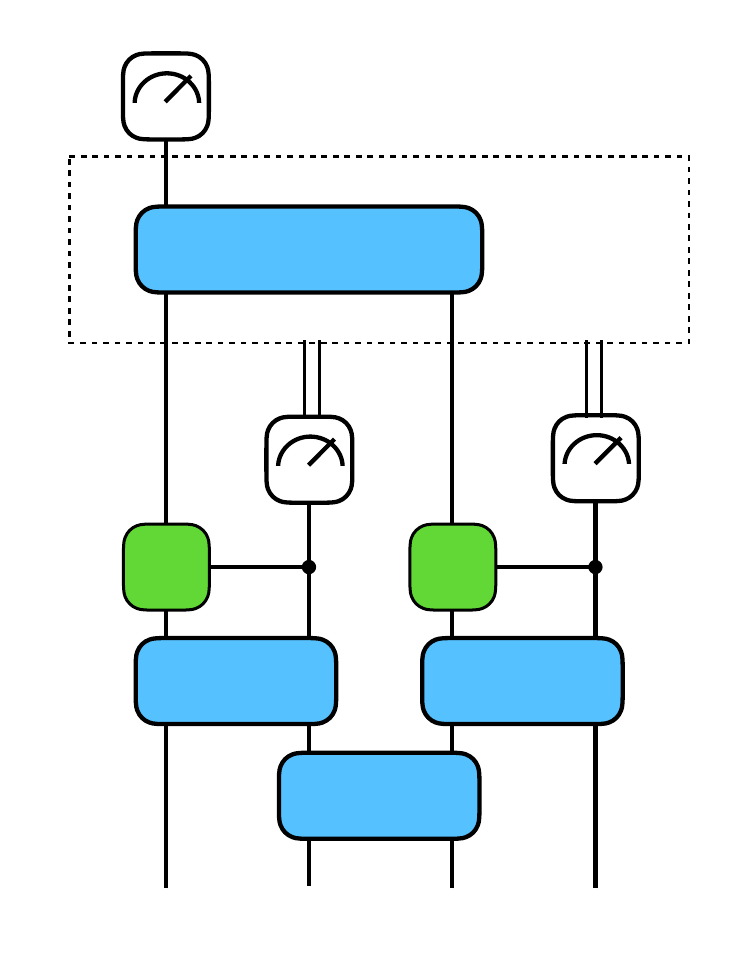}
    \caption{A four qubit bQCNN containing 111 trainable parameters (see Appendix \ref{struct} for details on gates). Note the dotted box connected to the measurements by classical double-lined rails. This dotted box indicates the presence of multiple (in this case four) distinct convolutional layers, which are chosen based on the mid-circuit measurement outcomes. The corresponding standard QCNN contains 66 parameters, and looks the same, though it lacks any mid-circuit measurements, and has only a single branch after the first pooling layer.}
    \label{fig:4qubit_bqcnn}
\end{figure}

\noindent As noted previously \cite{2019NatPh..15.1273C}, the pooling layers in QCNNs can be equivalently implemented by either controlled rotation gates or by mid-circuit measurements followed by classically conditioned operations on the remaining qubits. However, these operations only make use of local information. That is, a mid-circuit measurement of one qubit is used only to influence the rotation gate on a nearest-neighbor qubit. We would like to make use of the global measurement outcomes of all of the qubits that are measured at the pooling layer. This could allow for improved detection of long range correlations. We thus introduce the branching quantum convolutional neural network (bQCNN), which takes full advantage of the results of measurements at the pooling layers by conditioning the subsequent multi-qubit convolutional layers on the pooling layer measurement outcomes.
\newline
\newline
\noindent As a concrete example, consider a $N_{in}=4$ and $N_{out}=1$ QCNN. In the first convolutional layer, the four input qubits are entangled by parameterizable gates in both the QCNN and bQCNN. At the first pooling layer, we measure qubits 2 and 4. In the original QCNN architecture, a measurement result of 1 on qubit 2 would prompt the application of a parameterized rotation gate on qubit 1, and a 1 measurement on qubit 4 would yield a rotation on qubit 3. Measurements of 0 result in no conditioned rotations. In both approaches, qubits 1 and 3 then move on to the second convolutional layer, where more parameterized entangling gates are applied. In the bQCNN, however, we consider all four possible measurement outcomes of qubits 2 and 4. Each of the four outcomes --- 00, 01, 10, and 11 --- sends the remaining qubits 1 and 3 into one of four different convolutional layers --- different branches --- with distinct trainable parameters. Since only two qubits remain in the second convolutional layer, there is no need to apply a second pooling layer, and we can proceed directly to measurement of qubit 0, the classification qubit in both QCNN and bQCNN. In a larger circuit, however, we would repeat the convolution-pooling pattern until we reach a small number of output qubits. Even in this four qubit example, though, the bQCNN splits into four distinct branches, each with their own convolutional layer after the first pooling layer. The branching is denoted by the dashed box around the second convolutional layer in Fig. \ref{fig:4qubit_bqcnn}. The upshot of this is that the 4 qubit bQCNN contains 111 trainable parameters, while the equivalent QCNN contains only 66, despite both having the same overall circuit depth.
\newline
\newline
\noindent We have observed and rigorously evaluated that the classical branching structure of the bQCNN allows for a greater density of trainable parameters over a standard QCNN. Depending on the choice of specific circuit structure (depth of each convolutional layer and fraction of qubits to keep at each pooling layer), a bQCNN may have orders of magnitude more parameters than the QCNN of the same depth. For example, a 16 qubit bQCNN in which half of the qubits are discarded at each pooling layer, and which has only a single layer of entangling gates in its convolutional layers could have over 50,000 trainable parameters. Meanwhile its QCNN counterpart would have only about 400 parameters, with the same overall circuit depth. Of course, training 50,000 parameters with a classical optimizer is likely impractical, but we need not use every single branch in bQCNN in order to achieve a much more expressive ansatz than QCNN.
\newline
\newline
\noindent This increased density of parameters is potentially very useful in the NISQ era, as two qubit gate errors still present a significant limitation on the fidelity of deep circuits (for example, see \cite{2020arXiv200301293P} for recent data on error rates in Honeywell's trapped-ion device, which lists average two qubit gate error rates at $8\times 10^{-3}$). Again, we emphasize that this increase in parameter count is achieved with no additional circuit depth. If errors related to mid-circuit measurements are not large, bQCNNs can in principle express a greater range of operations for a given circuit depth compared to standard QCNNs, a point we explore in more detail in the next section.

\subsection{Expressibility of QCNN vs. bQCNN}
\label{exp_section}

\noindent The motivation for making use of information gained from mid-circuit measurements is to increase the expressibility of a parameterized, QCNN-like circuit without incurring significant error costs due increased circuit depth and gate count. That is, given an $N$-qubit parameterized hybrid circuit ansatz $\hat{U}(\boldsymbol{\theta})$, we seek to increase the volume of the full $N$-qubit Hilbert space accessible by applying $\hat{U}(\boldsymbol{\theta})$ --- for some set of parameters $\boldsymbol{\theta}$ --- to the all zero state, $|0\rangle^{\otimes N}$. To consider two extremes, a circuit $\hat{U}$ consisting of only single qubit gates will be unable to represent the majority of the Hilbert space, since it cannot generate entanglement from the all-zero product state. On the other hand, a circuit ansatz that can represent an arbitrary state in the Hilbert space would require up to $2^{2N}-1$ parameters (the dimension of $SU(2^N)$), and is therefore impractical for use as an ansatz in a hybrid quantum-classical algorithm beyond very small system sizes. 
\newline
\newline
\noindent Practical ans\"atze for hybrid algorithms lie somewhere between the exponentially-large, fully-parameterized circuit, and the trivial single qubit gate circuit in their expressibility. Here we show, using the specific notion of circuit expressibility introduced in \cite{2019arXiv190510876S}, that the bQCNN is more expressive than the standard QCNN of the same circuit depth. This, in principle, should allow for more complex classification tasks to be performed without increasing circuit depth and thus incurring significant errors from two-qubit gates. 
\newline
\newline
\noindent We emphasize that in an actual application of bQCNN we are not interested in creating states from the all-zero state. Rather, we are interested in disentangling input states in such a way as to induce a desired output probability distribution on the classification qubit. However, learning about the types of states that bQCNN and QCNN, respectively, can create is directly related to this question. If the circuit can create a particular state from the trivial all-zero state, then it is capable of disentangling that state to create a desired output. In order to maintain an entangled state after executing a full bQCNN and for ease of computation, we replace measurements and classical conditioning with multi-qubit controlled gates. However, this is equivalent to the mid-circuit measurement-based version of bQCNN from the perspective of the classification qubit, albeit with much greater circuit depth. (Alternatively, one could consider the space of mixed states to represent bQCNN, which would include all of the mid-circuit measurement outcomes and their associated branches).
\newline
\newline
\noindent We calculate the expressibility of a particular parameterized circuit ansatz $\hat{U}(\boldsymbol{\theta})$ by randomly generating many pairs of parameter sets, $\boldsymbol{\theta}, \boldsymbol{\phi}$, and then calculating the fidelity between the states generated by applying the corresponding parameterized circuits to the all zero state,

\begin{equation}
    F= |\langle 0| \hat{U}^\dagger(\boldsymbol{\theta}) \hat{U}(\boldsymbol{\phi})|0\rangle|^2.
\end{equation}

\noindent We sample over many random sets of parameters for both the bQCNN and the standard QCNN to obtain a probability distribution over the fidelities, $P_{(b)QCNN}(F)$.We then, using the Kullbeck-Liebler (KL) divergence, compare the sampled distributions to the Haar-random fidelity distribution,

\begin{equation}
    P_{Haar}(F)= (2^N-1)(F-1)^{2^N-2}.
\end{equation}

Computing the KL divergence of these two distributions yields a measure of the circuit ansatz in question, $\hat{U}(\boldsymbol{\theta})$, dubbed the expressibility:

\begin{equation}
    Expr= D_{KL}(P_{(b)QCNN}(F)||P_{Haar}(F))
    \label{expr_def}
\end{equation}

\begin{figure}
    \centering
    \includegraphics[width=8cm]{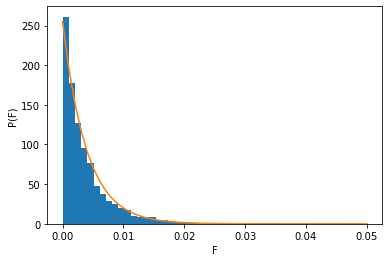}
    \includegraphics[width=8cm]{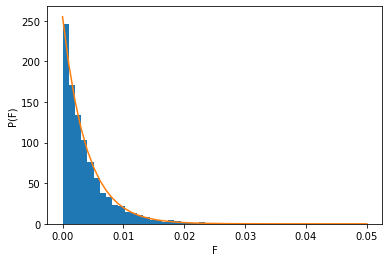}
    \caption{(Above) a histogram of fidelities for states generated by 8 qubit random QCNN circuits. (Below) The same for 8 qubit bQCNN circuits. In both cases the Haar fidelity distribution is superimposed as an orange line. The histograms visibly differ only slightly (the QCNN fidelities are more strongly peaked at 0), but the bQCNN is more than twice as expressive as the QCNN, according to the KL divergence. }
    \label{fig:fid_dist}
\end{figure}

\noindent The lower the value of \ref{expr_def} for a particular circuit, the more expressive that circuit is. In Fig. \ref{fig:fid_dist} we compare these distributions for 8 qubit QCNN and bQCNN circuits. We sample 4500 random pairs of circuits for each ansatz, and estimate the resulting expressibility (using 500 histogram bins) of the bQCNN to be 0.0072, while that of the standard QCNN is 0.0163, significantly higher, and therefore less expressive. Given that the 8 qubit bQCNN contains 914 trainable parameters, while the QCNN only contains 182, it is unsurprising that the bQCNN can more closely mimic the Haar distribution. This is despite requiring no additional circuit depth on a quantum device with mid-circuit measurement capabilities.
\newline
\newline
\noindent While this measure of expressibility by no means captures the full range of the differences between QCNN and bQCNN, it is an established metric that suggests that there are corners of the 4 qubit Hilbert space more easily accessible to bQCNN. This is of great potential benefit to QML tasks, especially considering the lack of increased circuit depth incurred.

\section{Artificial Training Tasks}

\begin{figure}
    \centering
    \includegraphics[width=8cm]{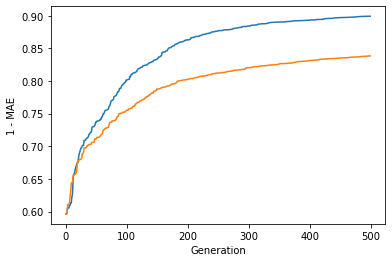}
    \caption{The correctness (one minus the mean absolute error) of the best circuit in a generation vs. training generation for a 4 qubit bQCNN (blue) and standard QCNN (orange) when training on bQCNN-generated data. To generate the data, a bQCNN circuit was created with a set of 111 random parameters. The complete set of 16 binary basis states ($|0000\rangle, |0001\rangle, |0010\rangle,$ etc.) was then passed through the circuit in reverse, and the resulting 16 states were used as training data. The binary label of the zeroth qubit before being passed through the inverted bQCNN, of course, served as the states' corresponding training labels. The above data was averaged over three training sessions on different randomly generated datasets. }
    \label{fig:bQCNN_on_bQCNN}
\end{figure}

\noindent To demonstrate that this added expressibility is of use, we can start by creating artificial training tasks. We can generate perfectly classifiable states using a 4 qubit bQCNN with random parameters by applying the inverse of (a particular branch of) the bQCNN to the complete set of product states in the computational basis (e.g. all 16 binary states for the 4 qubit circuit; $|0000\rangle, |0001\rangle, |0010\rangle,$ etc.), corresponding to every measurement outcome of the qubits after being passed through the forward circuit. If we use the zeroth qubit as the classification qubit, the span of the images of the set of states with $1$ or $0$ in the zeroth position under the inverted bQCNN corresponds to the two classes of states we are attempting to distinguish. Since these states are generated by running the bQCNN circuit with random parameters in reverse, a bQCNN of the same structure should in principle be able to learn to classify these randomly generated states without prior knowledge of the random parameters used to generate them. Indeed, this is what we find. In Fig. \ref{fig:bQCNN_on_bQCNN} we compare the training performance of a bQCNN and an ordinary QCNN on a set of 4 qubit bQCNN-generated training data. The bQCNN performs significantly better, with the correctness at 0.900 and 0.838 after 500 generations of training for the bQCNN and the QCNN, respectively.
\newline

\noindent We can consider an alternative demonstration of expressibility, in which we generate training data using a backwards QCNN, and then train both a QCNN and a bQCNN on this data. However, by setting all bQCNN branches to be equal to the single branch of the QCNN, we can trivially see that the bQCNN will always perform as well as QCNN, if not better, on a task that QCNN can in principle perform perfectly. Concerns about overfitting may arise when the bQCNN is used for tasks where QCNN performs very well, but we do not address out of training sample testing performance in this work, and leave this as a future question to pursue.

\if

We also consider the prospect of training a bQCNN on states generated by running an ordinary QCNN backwards. Here were create a standard 4 qubit QCNN by generating 66 random parameters and pass binary product states through the circuit backwards to create our set of 16 training states. The results of training a bQCNN and a QCNN on this data can be seen in Fig. \ref{fig:bQCNN_on_QCNN}, where the bQCNN demonstrates a correctness of 0.883, while the standard QCNN ends at 0.904 --- a much less significant difference than the two circuits exhibited when training on bQCNN-generated states. This is unsurprising, as the set of states one can generate by running a QCNN of a fixed depth backwards is a subset of the states one can generate using a bQCNN of the same depth. They are both MERA-type states \cite{2008PhRvL.101k0501V}, as pointed out in the case of QCNN in \cite{2019NatPh..15.1273C}. The multiple branches and resulting increased parameter count in the bQCNN, however, creates a greater variety of MERA-type states. When the branches differ from eachother by more than simple, single qubit gates, the resulting set of states may not be able to be parameterized by the corresponding QCNN, with its single, fixed set of convolutional layers.

\begin{figure}
    \centering
    \includegraphics[width=8cm]{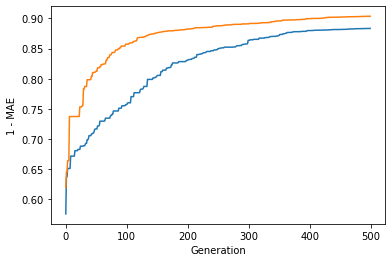}
    \caption{Training a 4 qubit, 111 parameter bQCNN (blue) and 66 parameter QCNN (orange) to classify a set of states generated by running a standard QCNN backwards. Here we see that, while initially slower to train, the bQCNN eventually catches up to the QCNN in performance. }
    \label{fig:bQCNN_on_QCNN}
\end{figure}

\fi

\section{Detecting an SPT Phase Transition}

\begin{figure}
    \centering
    \includegraphics[width=8cm]{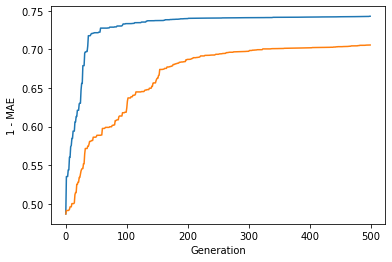}
    \caption{The training correctness (1 minus the mean absolute error cost function) for 500 generations of training of a 4 qubit bQCNN (blue) and 4 qubit QCNN on the diverse dataset. This training dataset features independent $X$ and $XXX$ perturbations in the Hamiltonian \ref{trainham} as a means of driving the phase transition. The relatively long-range perturbation introduced by the $XXX$ terms, as well as the diversity of the perturbed states in the training set allow for superior training performance in the bQCNN, which very rapidly approaches its ultimate maximum correctness of 0.743. Meanwhile the QCNN slowly reaches a final correctness value of 0.706, approximately five percent less than the bQCNN correctness, and after many more iterations of the genetic algorithm.}
    \label{fig:diverse_training}
\end{figure}

\noindent In earlier work \cite{2019NatPh..15.1273C}, the authors demonstrate the ability of a QCNN to recognize the phase transition between the $\mathbb{Z}_2\times \mathbb{Z}_2$ symmetry protected phase, which contains the 1D cluster state, and a trivial (either antiferromagnetic or paramagnetic) insulator phase. Specifically, the authors used as training data the ground states of the cluster Hamiltonian with a uniform magnetic field perturbation

\begin{equation}
    H= - \sum_i Z_{i-1}X_i Z_{i+1} - h_1 \sum_i X_i
\end{equation}

with varying values of the field strength, $h_1$. The trained QCNN was tested on the ground states of the same Hamiltonian, with the addition of an $XX$ Ising term,

\begin{equation}
    H= - \sum_i Z_{i-1}X_i Z_{i+1} - h_1 \sum_i X_i- h_2 \sum_i X_i X_{i+1},
\end{equation}

and was able to detect the phase transition for nonzero values of $h_2$, despite all of the training data having $h_2=0$. The QCNN excelled at this task because, for sufficiently small $h_1$ and $h_2$, the perturbations to the ground state induced by the magnetic field and the Ising term can be viewed as single or two site, local bit-flip errors applied to the cluster state. The QCNN pooling layer was able to learn to detect and correct these local errors, until they were strong enough to change the value of the string order parameters and result in a phase transition.
\newline
\newline
\noindent Towards diversifying the classification tasks well-suited to the bQCNN, we observe that the classical information obtained at the bQCNN pooling layers is global, and the decision of which branch to activate is based on the aggregate of measurement results of all of the discarded qubits. Thus, it is reasonable to conjecture that the bQCNN may be superior to QCNN at correcting non-local or multi-site errors. In addition, the presence of multiple branches suggests that the bQCNN might outperform QCNN when presented with a more diverse set of training data, with qualitatively different types of states within each class. Combining these two intuitions, we compare the training performance of bQCNN and QCNN in the phase recognition task using the ground states of the following Hamiltonian with open boundary conditions on 4 sites

\begin{equation}
    H= - \sum_i Z_{i-1}X_i Z_{i+1} - h \sum_i X_i- g \sum_i X_{i-1} X_i X_{i+1},
    \label{trainham}
\end{equation}

for various values of the field strengths, $h$ and $g$. Specifically, one half of the training dataset consists of the ground states of \ref{trainham} with $g=0, h \in [0,\frac{\pi}{2} )$, and the other consists of the ground states of \ref{trainham} with $h=0, g \in [0,\pi )$. To diagnose the phase transition, we use the simple string order parameter

\begin{equation}
    \hat{\mathcal{O}} = Z_1 Y_2 Y_3 Z_4,
\end{equation}

obtained by taking the dot product of the two $ZXZ$ stabilizer operators (for a larger system of $N$ qubits we can use $\hat{\mathcal{O}}= Z_1 Y_2 \left[ \prod_{i\in [3,N-2]} X_i \right] Y_{N-1} Z_N$). The expectation value of this order parameter is $1$ in the SPT phase and 0 in the trivial phase. Because of the small system size, the phase transition is smooth rather than abrupt, and the order parameter value decreases smoothly as we increase either $g$ or $h$, so we designate the point at which $\langle \mathcal{O} \rangle=0.5$ to be the phase transition. States with higher order parameter values are considered topological (denoted by a $0$ measurement at the end of the circuit), while those with lower values are trivial (denoted by a $1$ measurement). Using the ground states from \ref{trainham} and their associated phase labels, we can proceed to train the circuits.
\newline
\newline
\noindent We train a 4 qubit bQCNN containing 111 free parameters and a QCNN of the same width containing 66 parameters. The results of a 500 generation training session, in which we use a genetic algorithm (described in Appendix \ref{GA_app}) to minimize a mean absolute error cost function are depicted in Fig. \ref{fig:diverse_training}. 
\newline
\newline
\noindent Importantly, we note again that both circuits, when executed on a quantum device, have the same quantum circuit depth. Though the bQCNN contains mid-circuit measurements and immediate classical feedback --- a potential source of circuit error and latency --- a traditional QCNN constructed with sufficient circuit depth to achieve the same parameter count as the bQCNN would incur much larger two-qubit gate error costs for a sufficiently wide circuit. 

\section{Discussion and Conclusions} 

\noindent In this work we have introduced a new variational circuit ansatz for hybrid quantum machine learning applications. This ansatz, the branching QCNN, takes advantage of the emerging classical control flow capabilities of quantum devices --- that is, their ability to perform mid-circuit measurements and then decide which subsequent quantum operations to execute based on measurement results. Though the bQCNN has the same overall circuit depth as the original QCNN ansatz, it can contain many more trainable parameters, potentially increasing the range of QML tasks that can be performed on NISQ devices, where circuit depth is inherently limited by gate errors. We quantify the effect of increasing parameter count by numerically estimating the expressibility of an 8 qubit bQCNN, finding it to be significantly more expressive than its QCNN counterpart. Convinced of bQCNN's enhanced potential, we generate artificial training tasks in which a 4 qubit bQCNN demonstrates considerable advantage over QCNN. We then provide an example of a specific, physical training task in which bQCNN beats out QCNN, and speculate about the general class of training tasks for which this is true.
\newline
\newline
\noindent Beyond introducing the bQCNN ansatz, we emphasize the utility of mid-circuit measurement in near-term quantum applications. While we have used simple classical information processing to augment quantum operations --- namely, using a classical computer to decide which branch to execute after obtaining mid-circuit measurement results --- we envision more elaborate hybrid quantum-classical approaches in future. For example, we could imagine an approach in which classical information obtained from mid-circuit measurements is used to train classical machine learning methods, and the quantum and classical methods are used in conjunction to arrive at a result. Many other configurations of classical and quantum computing resources are conceivable. In any case, the ability to extract some classical information from a quantum algorithm via mid-circuit measurement should be tremendously useful in taking full advantage of near-term quantum devices. 
\newline
\newline
\noindent As a direct extension of this work, it would be useful to train and test bQCNN on larger systems, perhaps with the help of tensor network algorithms (e.g.\cite{2011AnPhy.326...96S}). Related to this, so-called ``barren plateaus'' occur in the cost functions of many variational ans\"atze as the parameter space grows large \cite{2018NatCo...9.4812M}, leading to questions about scalability of hybrid quantum-classical variational algorithms. There is some evidence to suggest that standard QCNNs avoid this issue \cite{2020arXiv201102966P}. It would be interesting to see if this is true for bQCNNs and other mid-circuit measurement-based hybrid algorithms.
\newline
\newline
\noindent A more detailed study of the training tasks for which bQCNN --- or other mid-circuit measurement-based ans\"atze --- are well-suited is warranted based on our results. In addition to the randomly-generated data, we presented one such specific example here, but it would be interesting to know more about the general classes of problems that one can and cannot solve by employing mid-circuit measurements. It would also be useful to know about the general class of MERA-like states generated by our random backwards circuits. In recent works, machine learning algorithms --- both quantum and classical --- have been used as ans\"atze for quantum states with interesting entanglement structure \cite{2017PhRvX...7b1021D}. 

\section{Outlook}
\noindent Further, there is a major opportunity in extending other ans\"atze for hybrid quantum algorithms with the use of mid-cicuit measurements. Some recent works have explored this direction. In particular, in \cite{2020arXiv200503023F}, the authors extend the length of a matrix product state ansatz via the use of mid-circuit measurement, and demonstrate the viability of this approach experimentally. Here we have modified a single such ansatz --- QCNN --- to make use of mid circuit measurements. This brings many other questions to the forefront: Are other circuit architectures amenable to similar modifications? Can we build hybrid variational ans\"atze from the ground-up using mid circuit measurements and classical control flow? Are there measurement-based quantum computing approaches to hybrid variational algorithms that could shed light on these questions? This last question has been explored in some recent works (e.g. \cite{2020arXiv201013940F}). In some sense, gate-based variational ans\"atze augmented by mid-circuit measurements and classical control flow can be viewed as a hybrid of a fully gate-based and a fully-measurement based protocol, if one views the the unitary gates preceding measurements as changes of measurement basis. For bQCNN and other variational circuits involving mid-circuit measurements to be of use on near-term devices, it is vital to understand their performance in realistic noisy environments. Today, only a small number of quantum devices offer mid-circuit measurement and classical feedback (e.g. \cite{2020arXiv200301293P}), and to the authors' knowledge, all of these devices are trapped-ion quantum computers. In the course of a mid-circuit measurement, scattered fluorescent photons from the detection process can result in bit-flips of far-away qubits, yielding a crosstalk error. Other errors can occur in the process of mid circuit measurement, and some amount of qubit idling occurs as classical information is processed after a measurement and before the subsequent branching. To understand the utility of bQCNN, we must know when the two qubit errors that would be incurred in a similarly expressive QCNN circuit (which will be deeper overall) outweigh the errors incurred by the measurements in a bQCNN. 
\newline
\newline
\noindent Going forward, as mid-circuit measurement is made available on more hardware, it will be productive to consider expanding the scope of the hybrid quantum-classical computation paradigm to make full use of existing quantum resources. This will mean allowing high-performance (soon exascale) classical computation to play a greater role in intermediate processing of data after mid-circuit measurements, in addition to executing classical optimization routines. The authors of \cite{2020arXiv200503023F} demonstrated that we can faithfully simulate a large quantum system using very few qubits by taking advantage of mid-circuit measurement. Some physical problems in condensed matter physics and materials science do not require the storage or manipulation of maximally-entangled, Page-like states; they often need only be entangled over a short range. This means that the combination of intermediate scale quantum devices and mid-circuit measurement capabilities could open the door to a new range of physical problems.

\begin{acknowledgments}
\noindent The authors thank Dr. Michael Fanto and Dr. Kathy-Anne (Brickman) Soderberg at the Air Force Research Laboratory (AFRL) Information Directorate for insightful discussions and comments on the work. We also thank Nidhi Aggarwal for useful comments and suggestions. This work is supported by Air Force STTR grant numbers FA8750-20-P-1721 and FA8750-20-P-1704. P.N. acknowledges support from the Army Research Office MURI (Ab-Initio Solid-State Quantum Materials) grant number W911NF-18-1-0431 (for supporting work on algorithms for
correlated quantum matter) and the DOE Office of Science, Basic Energy
Sciences (BES), Materials Sciences and Engineering Division
under FWP ERKCK47 (for supporting algorithms and computational approaches in highly entangled states and open quantum systems).
\end{acknowledgments}

\section{Appendix}
\label{struct}

\subsection{Circuit structure and implementation}

\begin{figure}

        \centering
        \includegraphics[width=.48\textwidth]{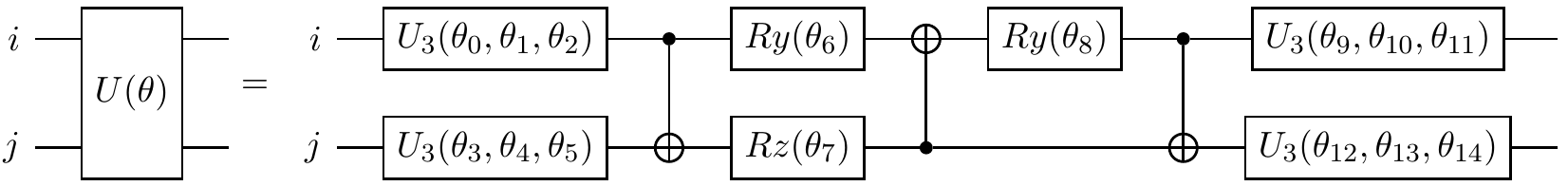}
        \caption{Gate decomposition for the general 2-qubit $SU(4)$ rotation applied to qubits $i$ and $j$ using cnot gates.}
        \label{fig:cnot_decomp}
\end{figure}

\begin{figure*}

        \includegraphics[width=\textwidth]{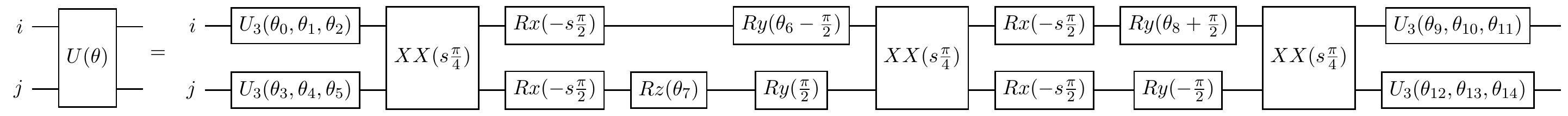}
        \caption{Gate decomposition for the general 2-qubit $SU(4)$ rotation applied to qubits $i$ and $j$ using trapped-ion native gates where $s=\pm{1}$ depending on the laser tuning of the device. }
        \label{fig:MS_decomp}

\end{figure*}

Any universal 2-qubit $SU(4)$ gate (depicted in blue in  \ref{fig:IM_vs_nonIM_circuit}) can be generally decomposed using at most 3 $CNOT$ gates \cite{Vatan_2004}, appropriately composed with single qubit rotations. For bQCNN simulations we utilize the gate decomposition shown in Fig. \ref{fig:cnot_decomp}, which would be appropriate for implementation on a hardware system with $CNOT$ as its native entangling gate (e.g. a superconducting qubit system). The input $\mathbf{\theta}$ is an array of fifteen rotation angles that fully parameterizes the two qubit rotation.

Equivalently, we could decompose an $SU(4)$ rotation into Molmer-Sorensen (MS) gates for implementation on a trapped-ion device as depicted in Fig. \ref{fig:MS_decomp}. Since trapped-ion devices are the only current providers of mid-circuit measurement capabilities, this decomposition would be more useful in a near-term implementation.

\subsection{Genetic Algorithm Parameter Optimization}
\label{GA_app}

Due to the variational nature of the QCNN, a proper training procedure for the parameterized circuit is necessary. We opt for a genetic algorithm rather than a gradient-based method due to the large number of parameters, and the poor scalability of gradient-based methods as this number continues to grow \cite{2018NatCo...9.4812M}. The genetic algorithm trains many generations of angle sets to slowly minimize a cost function. In our case, we use the following mean absolute error cost function:

\begin{equation}
    MAE= \sum_\alpha |  f(\alpha) - \langle f(\alpha) | \hat{U}(\boldsymbol{\theta})|\alpha \rangle |,
\end{equation}

where $\alpha$ is the label of a state $|\alpha\rangle$ in our training set, $f(\alpha)$ is the classification --- either 0 or 1 ---  of $|\alpha\rangle$, $\hat{U}(\boldsymbol{\theta})$ is our circuit ansatz (which is not unitary in the case of bQCNN, so this is an abuse of notation), and $|f(\alpha)\rangle$ is the state on qubit 0 that represents the correct classification outcome of $|\alpha\rangle$.

The genetic algorithm consists of the creation of successive generations, which successively improve based on the successes of the previous generations. Each generation consists of a population of parameter sets to fit the circuit. During each training cycle, the cost function is evaluated by running a predetermined number of shots and recording the classical measurement outcomes. Each rotation angle in each parameter set is then binarized into a single bitstring and arranged based on cost function performance. Some percentage of the best performing parameter sets immediately move on to the next generation and the remaining spots are filled through crossover between randomly chosen individuals, weighted by performance. 

Crossover is defined by the following: Given two parent bitstrings $a$ and $b$ of length $l$, they will produce two children bitstrings each of length l:
\begin{flalign*}
    & a_0a_1...a_l \xrightarrow{} a_0a_1...a_kb_{k+1}...b_l\\
    & b_0b_1...b_l \xrightarrow{} b_0b_1...b_ka_{k+1}...a_l
\end{flalign*}
where $k$ is the crossover point, chosen at random but only between parameters. Additionally, we specify a rate of mutation, which randomly swaps individual bits following crossover, introducing slight changes to angle values. The choice of these hyperparameters (population size, carry-over rate, mutation rate) to the genetic algorithm can significantly impact the training performance, so brief calibration of the hyperparameters was performed before any given training task.

\bibliography{main.bib}

\begin{thebibliography}{10}

\bibitem{Jordan255}
M.~I. Jordan and T.~M. Mitchell, ``Machine learning: Trends, perspectives, and
  prospects,'' {\em Science}, vol.~349, no.~6245, pp.~255--260, 2015.

\bibitem{2019ApPRv...6b1314B}
C.~D. {Bruzewicz}, J.~{Chiaverini}, R.~{McConnell}, and J.~M. {Sage},
  ``{Trapped-ion quantum computing: Progress and challenges},'' {\em Applied
  Physics Reviews}, vol.~6, p.~021314, June 2019.

\bibitem{2018arXiv180100862P}
J.~{Preskill}, ``{Quantum Computing in the NISQ era and beyond},'' {\em arXiv
  e-prints}, p.~arXiv:1801.00862, Jan. 2018.

\bibitem{2019arXiv190513641K}
M.~{Kjaergaard}, M.~E. {Schwartz}, J.~{Braum{\"u}ller}, P.~{Krantz}, J.~{I-Jan
  Wang}, S.~{Gustavsson}, and W.~D. {Oliver}, ``{Superconducting Qubits:
  Current State of Play},'' {\em arXiv e-prints}, p.~arXiv:1905.13641, May
  2019.

\bibitem{Dunjko_2018}
V.~Dunjko and H.~J. Briegel, ``Machine learning {\&} artificial intelligence in
  the quantum domain: a review of recent progress,'' {\em Reports on Progress
  in Physics}, vol.~81, p.~074001, jun 2018.

\bibitem{doi:10.1098/rspa.2017.0551}
C.~Ciliberto, M.~Herbster, A.~D. Ialongo, M.~Pontil, A.~Rocchetto, S.~Severini,
  and L.~Wossnig, ``Quantum machine learning: a classical perspective,'' {\em
  Proceedings of the Royal Society A: Mathematical, Physical and Engineering
  Sciences}, vol.~474, no.~2209, p.~20170551, 2018.

\bibitem{PhysRevLett.113.130503}
P.~Rebentrost, M.~Mohseni, and S.~Lloyd, ``Quantum support vector machine for
  big data classification,'' {\em Phys. Rev. Lett.}, vol.~113, p.~130503, Sep
  2014.

\bibitem{QML_review}
J.~Biamonte, P.~Wittek, N.~Pancotti, P.~Rebentrost, N.~Wiebe, and S.~Lloyd,
  ``Quantum machine learning,'' {\em Nature}, vol.~549, no.~7671, pp.~195--202,
  2017.

\bibitem{farhi2018classification}
E.~Farhi and H.~Neven, ``Classification with quantum neural networks on near
  term processors,'' 2018.

\bibitem{verdon2019quantum}
G.~Verdon, J.~Marks, S.~Nanda, S.~Leichenauer, and J.~Hidary, ``Quantum
  hamiltonian-based models and the variational quantum thermalizer algorithm,''
  2019.

\bibitem{verdonQGNN}
G.~Verdon, T.~McCourt, E.~Luzhnica, V.~Singh, S.~Leichenauer, and J.~Hidary,
  ``Quantum graph neural networks,'' 2019.

\bibitem{kim2017robust}
I.~H. Kim and B.~Swingle, ``Robust entanglement renormalization on a noisy
  quantum computer,'' 2017.

\bibitem{universal_quantum_control}
M.~Y. Niu, S.~Boixo, V.~N. Smelyanskiy, and H.~Neven, ``Universal quantum
  control through deep reinforcement learning,'' {\em npj Quantum Information},
  vol.~5, no.~1, p.~33, 2019.

\bibitem{2019QS&T....4b4001H}
W.~{Huggins}, P.~{Patil}, B.~{Mitchell}, K.~B. {Whaley}, and E.~M.
  {Stoudenmire}, ``{Towards quantum machine learning with tensor networks},''
  {\em Quantum Science and Technology}, vol.~4, p.~024001, Apr. 2019.

\bibitem{PhysRevLett.122.040504}
M.~Schuld and N.~Killoran, ``Quantum machine learning in feature hilbert
  spaces,'' {\em Phys. Rev. Lett.}, vol.~122, p.~040504, Feb 2019.

\bibitem{PhysRevA.101.032308}
M.~Schuld, A.~Bocharov, K.~M. Svore, and N.~Wiebe, ``Circuit-centric quantum
  classifiers,'' {\em Phys. Rev. A}, vol.~101, p.~032308, Mar 2020.

\bibitem{2019Natur.567..209H}
V.~{Havl{\'\i}{\v{c}}ek}, A.~D. {C{\'o}rcoles}, K.~{Temme}, A.~W. {Harrow},
  A.~{Kandala}, J.~M. {Chow}, and J.~M. {Gambetta}, ``{Supervised learning with
  quantum-enhanced feature spaces},'' {\em \nat}, vol.~567, pp.~209--212, Mar.
  2019.

\bibitem{Wan_2017}
K.~H. Wan, O.~Dahlsten, H.~Kristjánsson, R.~Gardner, and M.~S. Kim, ``Quantum
  generalisation of feedforward neural networks,'' {\em npj Quantum
  Information}, vol.~3, Sep 2017.

\bibitem{Romero_2017}
J.~Romero, J.~P. Olson, and A.~Aspuru-Guzik, ``Quantum autoencoders for
  efficient compression of quantum data,'' {\em Quantum Science and
  Technology}, vol.~2, p.~045001, aug 2017.

\bibitem{Carleo602}
G.~Carleo and M.~Troyer, ``Solving the quantum many-body problem with
  artificial neural networks,'' {\em Science}, vol.~355, no.~6325,
  pp.~602--606, 2017.

\bibitem{2017arXiv170605413O}
J.~{Olson}, Y.~{Cao}, J.~{Romero}, P.~{Johnson}, P.-L. {Dallaire-Demers},
  N.~{Sawaya}, P.~{Narang}, I.~{Kivlichan}, M.~{Wasielewski}, and
  A.~{Aspuru-Guzik}, ``{Quantum Information and Computation for Chemistry},''
  {\em arXiv e-prints}, p.~arXiv:1706.05413, June 2017.

\bibitem{butler2018machine}
K.~T. Butler, D.~W. Davies, H.~Cartwright, O.~Isayev, and A.~Walsh, ``Machine
  learning for molecular and materials science,'' {\em Nature}, vol.~559,
  no.~7715, pp.~547--555, 2018.

\bibitem{head2020quantum}
K.~Head-Marsden, J.~Flick, C.~J. Ciccarino, and P.~Narang, ``Quantum
  information and algorithms for correlated quantum matter,'' {\em Chemical
  Reviews}.

\bibitem{2016arXiv160505775M}
E.~{Miles Stoudenmire} and D.~J. {Schwab}, ``{Supervised Learning with
  Quantum-Inspired Tensor Networks},'' {\em arXiv e-prints},
  p.~arXiv:1605.05775, May 2016.

\bibitem{2016NJPh...18b3023M}
J.~R. {McClean}, J.~{Romero}, R.~{Babbush}, and A.~{Aspuru-Guzik}, ``{The
  theory of variational hybrid quantum-classical algorithms},'' {\em New
  Journal of Physics}, vol.~18, p.~023023, Feb. 2016.

\bibitem{farhi2017quantum}
E.~Farhi, J.~Goldstone, S.~Gutmann, and H.~Neven, ``Quantum algorithms for
  fixed qubit architectures,'' 2017.

\bibitem{2019NatPh..15.1273C}
I.~{Cong}, S.~{Choi}, and M.~D. {Lukin}, ``{Quantum convolutional neural
  networks},'' {\em Nature Physics}, vol.~15, pp.~1273--1278, Aug. 2019.

\bibitem{franken2020explorations}
L.~Franken and B.~Georgiev, ``Explorations in quantum neural networks with
  intermediate measurements,'' in {\em Proceedings of ESANN}, 2020.

\bibitem{2017NatSR...7.8823B}
P.~{Broecker}, J.~{Carrasquilla}, R.~G. {Melko}, and S.~{Trebst}, ``{Machine
  learning quantum phases of matter beyond the fermion sign problem},'' {\em
  Scientific Reports}, vol.~7, p.~8823, Aug. 2017.

\bibitem{2020arXiv200301293P}
J.~M. {Pino}, J.~M. {Dreiling}, C.~{Figgatt}, J.~P. {Gaebler}, S.~A. {Moses},
  M.~S. {Allman}, C.~H. {Baldwin}, M.~{Foss-Feig}, D.~{Hayes}, K.~{Mayer},
  C.~{Ryan-Anderson}, and B.~{Neyenhuis}, ``{Demonstration of the QCCD
  trapped-ion quantum computer architecture},'' {\em arXiv e-prints},
  p.~arXiv:2003.01293, Mar. 2020.

\bibitem{2020arXiv200603071E}
A.~{Erhard}, H.~{Poulsen Nautrup}, M.~{Meth}, L.~{Postler}, R.~{Stricker},
  M.~{Ringbauer}, P.~{Schindler}, H.~J. {Briegel}, R.~{Blatt}, N.~{Friis}, and
  T.~{Monz}, ``{Entangling logical qubits with lattice surgery},'' {\em arXiv
  e-prints}, p.~arXiv:2006.03071, June 2020.

\bibitem{ibm_quantum_experience_2020}
``Docs and resources,'' 2020.

\bibitem{2009arXiv0910.1116B}
H.~J. {Briegel}, D.~E. {Browne}, W.~{D{\"u}r}, R.~{Raussendorf}, and M.~{Van
  den Nest}, ``{Measurement-based quantum computation},'' {\em arXiv e-prints},
  p.~arXiv:0910.1116, Oct. 2009.

\bibitem{lecun2015deep}
Y.~LeCun, Y.~Bengio, and G.~Hinton, ``Deep learning,'' {\em nature}, vol.~521,
  no.~7553, pp.~436--444, 2015.

\bibitem{2019arXiv190510876S}
S.~{Sim}, P.~D. {Johnson}, and A.~{Aspuru-Guzik}, ``{Expressibility and
  entangling capability of parameterized quantum circuits for hybrid
  quantum-classical algorithms},'' {\em arXiv e-prints}, p.~arXiv:1905.10876,
  May 2019.

\bibitem{2011AnPhy.326...96S}
U.~{Schollw{\"o}ck}, ``{The density-matrix renormalization group in the age of
  matrix product states},'' {\em Annals of Physics}, vol.~326, pp.~96--192,
  Jan. 2011.

\bibitem{2018NatCo...9.4812M}
J.~R. {McClean}, S.~{Boixo}, V.~N. {Smelyanskiy}, R.~{Babbush}, and H.~{Neven},
  ``{Barren plateaus in quantum neural network training landscapes},'' {\em
  Nature Communications}, vol.~9, p.~4812, Nov. 2018.

\bibitem{2020arXiv201102966P}
A.~{Pesah}, M.~{Cerezo}, S.~{Wang}, T.~{Volkoff}, A.~T. {Sornborger}, and P.~J.
  {Coles}, ``{Absence of Barren Plateaus in Quantum Convolutional Neural
  Networks},'' {\em arXiv e-prints}, p.~arXiv:2011.02966, Nov. 2020.

\bibitem{2017PhRvX...7b1021D}
D.-L. {Deng}, X.~{Li}, and S.~{Das Sarma}, ``{Quantum Entanglement in Neural
  Network States},'' {\em Physical Review X}, vol.~7, p.~021021, Apr. 2017.

\bibitem{2020arXiv200503023F}
M.~{Foss-Feig}, D.~{Hayes}, J.~M. {Dreiling}, C.~{Figgatt}, J.~P. {Gaebler},
  S.~A. {Moses}, J.~M. {Pino}, and A.~C. {Potter}, ``{Holographic quantum
  algorithms for simulating correlated spin systems},'' {\em arXiv e-prints},
  p.~arXiv:2005.03023, May 2020.

\bibitem{2020arXiv201013940F}
R.~R. {Ferguson}, L.~{Dellantonio}, K.~{Jansen}, A.~{Al Balushi}, W.~{D{\"u}r},
  and C.~A. {Muschik}, ``{A measurement-based variational quantum
  eigensolver},'' {\em arXiv e-prints}, p.~arXiv:2010.13940, Oct. 2020.

\bibitem{Vatan_2004}
F.~Vatan and C.~Williams, ``Optimal quantum circuits for general two-qubit
  gates,'' {\em Physical Review A}, vol.~69, Mar 2004.

\end{thebibliography}
\bibliographystyle{ieeetr}

\end{document}